\documentclass[aps, pre, preprint, groupedaddress, amsmath, amssymb]{revtex4-1}
\usepackage{graphicx}  % Include figure files
\usepackage{dcolumn}   % Align table columns on decimal point
\usepackage{bm}        % bold math
\usepackage{multirow}
\begin{document}

\title{Image-based Analysis of Patterns Formed in Drying Drops}

\author{Anusuya Pal$^{a}$, Amalesh Gope$^{b}$ and Germano S. Iannacchione$^{a}$} \email{gsiannac@wpi.edu}

\affiliation{$^{a}$ Order-Disorder Phenomena Laboratory, Department of Physics, Worcester Polytechnic Institute, Worcester, MA, 01609, USA \\ $^{b}$ Department of English and Foreign Languages, Tezpur University, Tezpur, Assam, 784028, India}

%\date{\today}

%***********************************************************************

\begin{abstract}
Image processing and pattern recognition offer a useful and versatile method for optically characterizing drops of a colloidal solution during the drying process and in its final state. This paper exploits image processing techniques applied to cross-polarizing microscopy to probe birefringence and the bright-field microscopy to examine the morphological patterns. The bio-colloidal solution of interest is a mixture of water, liquid crystal (LC) and three different proteins [lysozyme (Lys), myoglobin (Myo), and bovine serum albumin (BSA)], all at a fixed relative concentration. During the drying process, the LC phase separates and becomes optically active detectable through its birefringence. Further, as the protein concentrates, it forms cracks under strain due to the evaporation of water. The mean intensity profile of the drying process is examined using an automated image processing technique that reveals three unique regimes- a steady upsurge, a speedy rise, and an eventual saturation. The high values of standard deviation show the complexity, the roughness, and inhomogeneity of the image surface. A semi-automated image processing technique is proposed to quantify the distance between the consecutive cracks by converting those into high contrast images. The outcome of the image analysis correlates with the initial state of the mixture, the nature of the proteins, and the mechanical response of the final patterns. The paper reveals new insights on the self-assembly of the macromolecules during the drying mechanism of any aqueous solution. 

\end{abstract}


\maketitle

\section{Introduction}

Image processing and automated pattern recognition techniques are crucial elements in advance of (optical) imaging technology. As image acquisition has become more precise (in time and spatial resolution) and ubiquitous, the need to extract information from these images has grown exponentially. One of the many scientific areas impacted by such processing is in observing the drying drops of some complex fluids. Specifically, one such class of complex colloidal fluid relevant to forensics and medical sciences is biological fluid such as blood, blood serum, tear, urine, or simple protein solutions \cite{brutin2018recent}. Recent research findings \cite{chen2016blood,bel2019morphology} have found that dried drops of serums can differentiate the state of health of a patient. Furthermore, the whole dried pattern appears to be connected to the initial state of the material present in these complex fluids \cite{pal2019comparative,pal2019phase}. For example, Deegan et al. have studied colloidal drops and found that the drops get pinned after depositing on the substrate. The particles start concentrating near the edge more than the central region, forming a ring called ``coffee-ring effect" \cite{deegan1997capillary}. Due to this pinning and solvent loss, the drops are no longer able to shrink and ultimately, form cracks (patterns) under strain during the drying process. 

This paper exploits image processing techniques applied to the cross-polarizing microscopy to probe birefringence and the bright-field microscopy to examine the crack patterns. The bio-colloidal solution of interest is a mixture of water, and three different proteins [lysozyme (Lys), myoglobin (Myo), and bovine serum albumin (BSA)], all at a fixed relative concentration with and without a nematic liquid crystal (LC). The patterns are characterized using three different criteria, (a) the optical outcome, (b) the textural dynamics, and, (c) statistical significance. This work uses both automated and semi-automated image processing techniques to quantify drying dynamics and dried morphology. Our analysis reveals that the topmost part of the BSA film is coated with LC droplets, whereas, the same LC droplets are randomly distributed underneath the Lys and Myo films. This paper demonstrates the versatility and usefulness of these image processing techniques and is adaptable to any similar imaging system.

\section{Materials, Sample Preparation and Image Acquisition} 
Lysozyme (Lys), myoglobin (Myo) and bovine serum albumin (BSA) are globular proteins differing in mass, charge, and properties \cite{pal2019comparative}. The presence of a heme group helps Myo to serve as an oxygen-binding protein. Lys and Myo have similar molecular masses of $ \sim14.3$, and $ \sim17.0$~KDa respectively, whereas BSA contains a higher molecular mass of $ \sim66.5$~kDa \cite{pal2019comparative}. The net charge of Lys, BSA, and Myo is positive, neutral, and negative respectively. The commercial powdered proteins were purchased from Sigma Aldrich, USA. Myo (Catalog no. M0630), Lys (L6876), and BSA (A2153) were used without any further purification. Distinct protein solutions containing each of Lys, Myo, and BSA ($100$~mg each) were prepared by dissolving those separately in $1$~mL of de-ionized water. Following that, an optically active and polar LC, 5CB (328510) was heated above the transition temperature ($35$~$^{\circ}$C), and $10$~$\mu$L was added as a third component to the solution. A $\sim1.3$~$\mu$L of the solution was pipetted on the coverslip in the form of a circular drop, which was allowed to dry under ambient conditions \cite{pal2019comparative}. 

\begin{figure}[h]
\includegraphics[width=\textwidth]{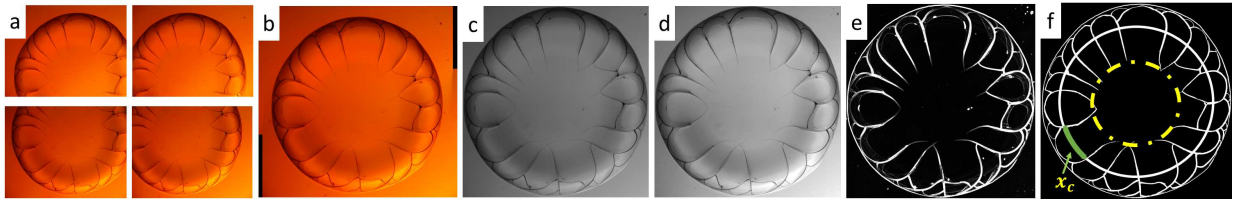}
\caption{A representative processing of images of the dried state of BSA drop to quantify the spacing between the cracks. a) The raw microscopic images. b) A stitched image. c) An $8$-bit gray image. d) An adjusted brightness-contrast image. e) A high contrast image after filtering through variance filter. f) A final binary image depicting different regions, circular line-cut, and crack spacing ($x_c$).} \label{fig1}
\end{figure}

A polarizing microscopy (Leitz Wetzlar, Germany) with a $5\times$ objective lens at a fixed resolution of $2048 \times 1536$~pixels was used to capture the images \cite{pal2019comparative,pal2019phase}. An $8$-bit digital camera (Amscope MU300) in color profile mode assigns intensity values from $0$ to $255$ in each color of red, green, and blue (RGB) pixels for image acquisition. After the deposition of the drop on the coverslip, the time-lapse images in crossed polarizing configuration were recorded at every two seconds to monitor LC droplets' birefringence during the drying process. Each drop was deposited three times from the same sample to ensure the reproducibility. The final dried state was examined in bright-field configuration to understand different crack patterns in different protein drops with and without inclusion of LC droplets. Since it was challenging to view the entire $\sim2$~mm diameter drop using a $5\times$ objective lens, different sections of the dried drops were imaged separately (Fig.~\ref{fig1}a). Each section is stitched together (Fig.~\ref{fig1}b) by using \textit{Stitching} plugin \cite{preibisch2009globally} in Fiji \cite{schindelin2012fiji}. We propose a semi-automated image-processing algorithm on the stitched images to determine the spacing between the consecutive cracks ($x_c$) in ImageJ \cite{abramoff2004image}. We converted stitched bright-field images into gray (Fig.~\ref{fig1}c). For these gray images, the range of monochromatic shades from black to white are displayed from $0$ to $255$ without partitioning into RGB sets of pixels. Then, the bright and contrast of the image were adjusted (Fig.~\ref{fig1}d) and filtered with a variance filter (Fig.~\ref{fig1}e), and processed into an 8-bit binary image ($255$~for pixels depicting the crack lines and $0$~for pixels elsewhere else) (Fig.~\ref{fig1}f) \cite{pal2019comparative,pal2019phase}. However, the filtering was not enough to process the images into exact binary images. The artifacts were, therefore manually removed by comparing (overlapping) the processed images on to the original images. However, future work will automate this procedure. 

The morphology of the dried drop is such that it can be divided into two regions: rim and central. The rim is near the edge depicted by yellow dashed line in Fig.~\ref{fig1}f. We converted all the images of the dried drops into a scaled stack. Three circular-cut lines (shown by white circular line in Fig.~\ref{fig1}f) of different radii were made in each region by using \textit{Oval Profile} plugin in ImageJ \cite{pal2019comparative,pal2019phase}. The intensity values were plotted along each circular line at every $0.1^\circ$ as a function of arc-length along each circle. A script using `\textit{Array.findMaxima}' was used to determine the positions of maximum intensity values. An estimate of the crack spacing, $x_c$ (outlined by green color in Fig.~\ref{fig1}f) is calculated by the consecutive maxima difference \cite{pal2019comparative,pal2019phase}. All the intensity values along the crack lines were cross-checked manually to see if there is an artifact. If found, were corrected to the consistent (and fixed) values to maintain uniformity along the crack lines. A threshold of $\pm0.005$~mm were used as a standard range and $x_c $ values outside this range was recorded and aggregated in each region to obtain an average ($\overline{x}_c$) \cite{pal2019comparative,pal2019phase}. However, there may be cracks that propagate along the drawn circle, leading to some uncertainty in the extracted crack distributions.

To quantify the birefringence of the LC droplets during the drying process, a circular region of interest (ROI) was drawn on the image covering the area in the drop, and another ROI was drawn on the image covering the background (coverslip) using the \textit{Oval tool} in ImageJ \cite{abramoff2004image}. Once the ROI was selected, the birefringence intensity of that ROI was measured. To ensure that the different size of ROI in the sample and the background do not affect the intensity measurement, \textit{mean gray value} (the sum of the values of the pixels in the selection divided by the area of the pixels in the selection) in ImageJ \cite{abramoff2004image} was chosen. A script was written for an automated image-processing algorithm which measures the \textit{mean gray value} for each ROI in the images during the drying process. However, the differences in the \textit{mean gray values} during image recording might affect both the background and the sample. To counter this, a correction (calibration) with the background i.e., a correction factor was determined. It could be done by subtracting the background gray values; however, it wouldn't have fixed an uneven background in the series of images. So, the ROI with the lowest background \textit{mean gray value} was chosen from the whole set of images during the drying process as a reference. The lowest was also selected to avoid into running the risk of generating overexposed images (otherwise the correction factors would be smaller than $1$). The corrected intensity of the sample ($I_c$) was then determined for each sample by dividing the mean gray values with the correction factor. The intensity was averaged over a range of $30$ seconds and also averaged for three drops. Good reproducibility was found in the intensity profile, though the time was shifted (added or subtracted) to make the profile nearly overlap to each other. Note that the lamp intensity was kept fixed throughout the whole experiment done with each protein drop. And finally, the averaged corrected intensity values ($\overline{I}_c$) was plotted with time. In addition to $\overline{I}_c$, another textural parameter i.e., standard deviation (SD) was calculated and were averaged over $30$ seconds to understand the emerging complexity during the drying process. 

\section{Results and Discussions}

\begin{figure}[h]
\includegraphics[width=\textwidth]{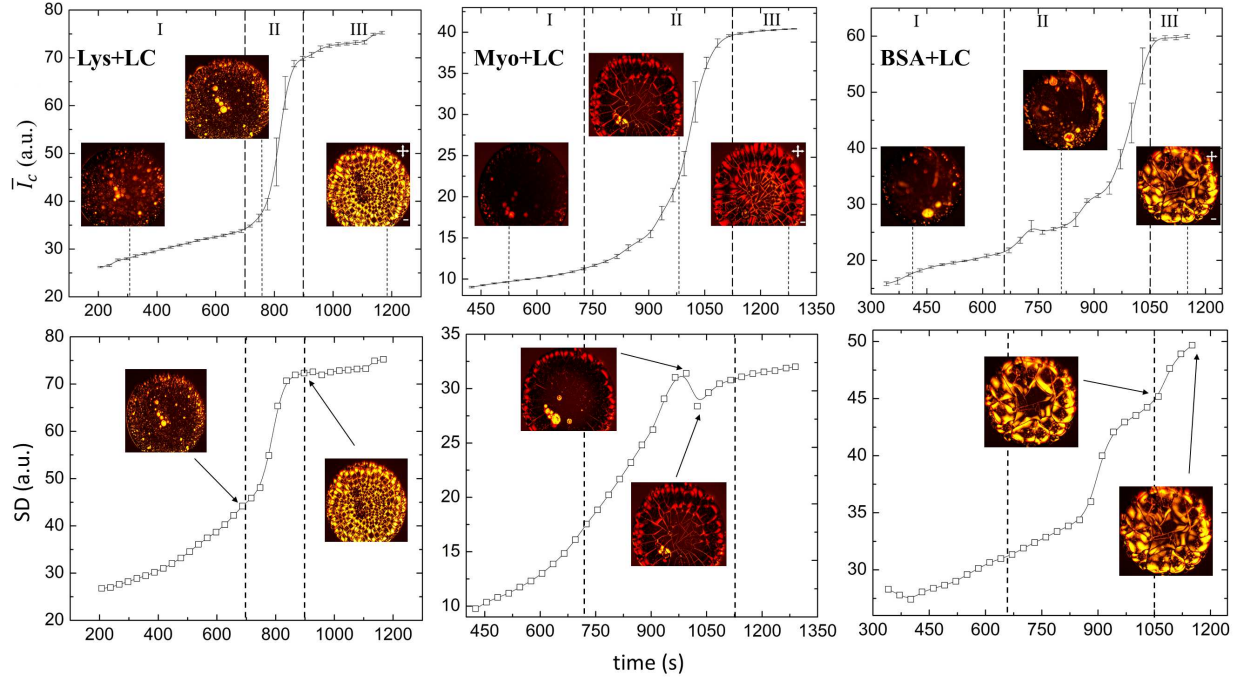}
\caption{Upper panel: Variation of birefringence intensity ($\overline{I}_c$) of LC droplets during the drying process. Three unique regimes are observed: (I) a steady upsurge, (II) a speedy rise, and, (III) an eventual saturation of intensity. Lower panel: Variation of standard deviation (SD) of the images during different drying stages.} \label{fig2}
\end{figure}

The upper panel of Fig. 2 reveals the variation of birefringence intensity ($\overline{I}_c$) of LC droplets during the drying process. The standard deviation (n = 3) is higher in all the transition compared to the other regimes due to the presence of different sized LC droplets in different drops. It demonstrates an archetypal response of LC droplets dissolved in protein solutions and displays three distinct regimes: (I) the initial process of getting a minimum contact angle configuration with a gradual increase of intensity, (II) the intensity gets amplified due to LC activities, and (III) the intensity is saturated in the final regime once the drop completes the evaporation process. It also suggests that the birefringence nature is independent of types of proteins and less efficient to capture the minute details of the LC activities during the drying process. The stage I completes in $11$-$14$~minutes, and the rate of intensity in stage II is $\sim0.28$ (for Lys), $\sim0.09$ (for Myo), and $\sim0.08$ (for BSA). A smooth transition regime in both Lys and Myo is observed, but not in BSA drops indicating a distinct final morphology in BSA. As mentioned earlier, BSA contains a higher molecular weight ($\sim66.5$~KDa) and the increase of mass in proteins decreases the rate of intensity change (one surprising finding is the drastic decrease from $0.28$ to $0.09$ in similar weighted proteins of Lys and Myo respectively). It is to note that LC has a strong dipole moment (polar) and the presence of a heme group without any disulfide bonds in Myo could interrupt the local interactions of Myo-LC differently than Lys-LC. Moreover, LC droplets wet the substrate (coverslip) differently for different proteins because of their difference in nature (mass, size, charges of proteins). Hence, it indicates that the mass is not only the contributing factor but involves many other interactions such as self-assembly among protein particles (protein-protein interaction), phase separation and self-assembly of LC droplets (protein-LC, LC-substrate), etc. \cite{pal2019comparative}.

The lower panel of Fig.~\ref{fig2} shows the SD (standard deviation of the image) of the same drops depicting a similar trend as a function of time. The $\overline{I}_c$ captures the whole process such as phase separation, formation of protein film, movement of fluid, water evaporation, etc., whereas, the SD is sensitive to changes in the image complexity \cite{carreon2018texture}. The SD gives additional insights into the morphology in terms of changes in inhomogeneity and complexity. In Myo, it shows a difference in the evolution pattern from $\overline{I}_c$ by observing a dip and rise in the transition region. Though Myo protein film is faint red in color under cross-polarizing microscopy, it doesn't mean that it is optically active like the LC droplets (yellow in color). It means that Myo reflects a particular wavelength (red), and absorbs rest of the colors in the visible spectrum. This becomes advantageous over other proteins in our study i.e., Lys and BSA because the inhomogeneity is clearly distinguishable by calculating SD. The flow of the LC droplets through the Myo film promotes a dip, and the process of turning the film (into reddish) indicates a difference in SD. On the other hand, BSA shows a rise in the complexity of regime III. At this stage, the texture (SD) is found to be changing while distributing the LC droplets whereas the $\overline{I}_c$ fails to display such details and shows constant values. In spite of these differences, a common characteristic is observed in both $\overline{I}_c$ and SD by their increase through the drying process and saturation when the drops are fully dried. This is consistent with the process of droplet drying, phase separation, film formation, stain, and finally, film cracking.  

\begin{figure}
\includegraphics[width=\textwidth]{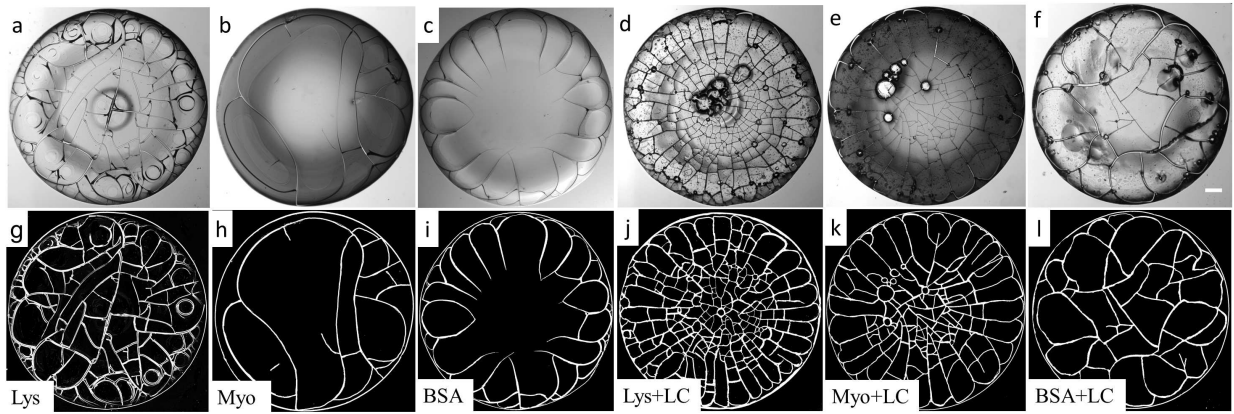}
\caption{Upper panel: After 24 hours, the images were captured in the bright-field configuration in the presence and absence of LC droplets (a-f). Lower panel: the processed binary images (g-l). The scale bar of $0.15$~mm is shown in the right corner of [f].} \label{fig3}
\end{figure}

Fig.~\ref{fig3} shows the crack patterns and their processed binary images (in presence and absence of LC droplets). The common feature noticed in all the drops is the presence of ``coffee-ring" \cite{deegan1997capillary} (upper panel of Fig.~\ref{fig3}) confirming the general mechanism of the drying colloidal drops. The inclusion of the LC droplets affects the cracks' nature, and makes a rough protein film's surface. LC is believed to increase the stress in all the drops and the stress in turns produces more number of cracks. The central region of both Myo and BSA does not contain any cracks in the absence of the LC droplets (Fig.~\ref{fig3}b,c). Contrary to that, the inclusion of the LC droplets in the protein drops leads to- (i) an ordered crack distribution and disturbs a `mound'-like structure in the Lys drops (Fig.~\ref{fig3}a,d), (ii) a formation of well-connected large and small crack domains in the Myo (Fig.~\ref{fig3}b,e), and (iii) an overall increase of the cracks in the BSA and the formation of a few cracks in the central region (Fig.~\ref{fig3}c,f).

\begin{figure}[h]
\includegraphics[width=\textwidth]{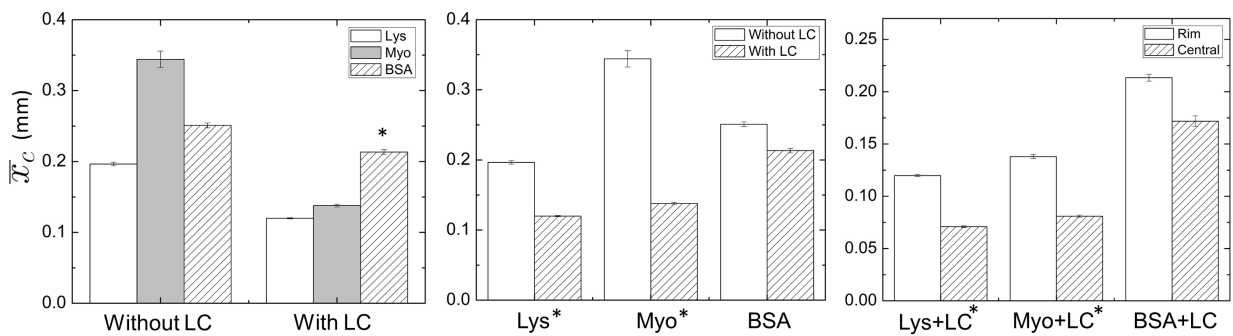}
\caption{The comparison of mean crack spacing ($ \overline{x}_c$) in different conditions and environment. Left panel: comparison of $ \overline{x}_c$ with and without LC droplets observed in the rim region. Middle panel: overall effect of the LC droplets in relation to $ \overline{x}_c$ in all the protein drops. Right panel: comparison of $ \overline{x}_c$ in the rim and the central region in different protein drops in presence of LC droplets. An asterisk [*] indicates significant pairs.} \label{fig4}
\end{figure}

To quantify the crack spacing ($x_c$) and to develop a physical mechanism of the underlying process of crack distributions, a non-parametric Mann-Whitney U and Kruskal Wallis tests were conducted. Any statistical difference in mean crack spacing ($\overline{x}_c$) is considered to be significant when p $ <0.05$. The significant pairs are marked with an asterisk [*] in Fig.~\ref{fig4}. All the experimental data are expressed as mean $\pm$ standard error (SE). The crack morphology discussed in Fig.~\ref{fig3} confirmed that the rim region of all the protein drops is specified for more number of cracks compared to the central region. We, therefore, conducted a series of statistical tests to examine the (significant) effects of LC droplets on all the proteins. The left panel of Fig.~\ref{fig4} shows the way LC droplets affect the $x_c$ distributions in the rim region of all proteins. We did not observe any significant difference related to $x_c$ distributions. The distribution is observed to be equally large in all the protein solutions without LC droplets. However, $x_c$ is observed to be reduced (i.e., there is more number of cracks) when LC droplets are added. A pairwise comparison (using Dunn’s procedure with a Bonferroni correction) of the proteins in the presence of LC droplets reveals that BSA still has significantly larger $x_c$ compared to the other two protein drops (Lys and Myo). In other words, LC droplets do not have a significant effect in terms of $x_c$ distributions on BSA even though the morphology shows a reduced $x_c$ when LC is added (Fig.~\ref{fig3}f,l). The middle panel of Fig.~\ref{fig4} further reveals that the presence of LC considerably reduces the $x_c$ distributions (a large number of uniformed cracks) in all the proteins, however, the difference is found to be statistically significant in both Lys and Myo, but not in the BSA. Finally, we examined the way (inclusion of) LC droplets affect the $x_c$ distributions in the rim and the central regions of all the protein drops (right panel of Fig.~\ref{fig4}). The results confirm that the presence of LC considerably reduces the $x_c$ in the central region of all the proteins. However, this reduction of $x_c$ is observed to be significant in both Lys and Myo (but not in BSA). This outcome is evident as we have already seen that LC does not affect the BSA (in terms of $x_c$ reduction). The quantification of $x_c$ and the outcome of the statistical tests provide a convincing indication about the effect of LC on the proteins (under study) which otherwise is slightly misleading if the morphological observations are considered alone. 

Following these findings discussed above, a physical mechanism is proposed. Our findings confirm that Lys and Myo follow a similar process of crack formation, unlike BSA. Each domain in these two proteins is buckled up, and most of the LC droplets are moved underneath the film from the cracks through the capillary action. The dark region at the center of each of these regions is the film attached to the coverslip. In contrast, BSA shows a radial defect (four-fan brushes) which is commonly observed at the LC-air interface confirming that the LC droplets are spread at the top of the BSA film \cite{pal2019comparative}.

\section{Conclusions}
Sessile drop drying is popularly used to explore the self-assembly and the phase separation of the particles in the colloidal drops. In this paper, we proposed an automated image-based mechanism to calculate birefringence of the optically active material and can be applied to any such colloidal drops. Further, the proposed semi-automated image processing mechanism of quantifying crack patterns can be applied to any drying system (with similar crack patterns) helping to understand the underlying physical mechanism. These methods have a broad range of applications quantifying image morphology in an efficient and economical way. 

% REFERENCES:

\bibliography{bookch_premi}
\end{document}